\documentclass[prb,twocolumn]{revtex4}   

\usepackage{graphicx}
\usepackage{color}

\begin{document}

\title
{Supersolid state in fermionic optical lattice systems}

\author
{
Akihisa Koga,$^1$ Takuji Higashiyama,$^2$ Kensuke Inaba,$^2$
Seiichiro Suga,$^2$ and Norio Kawakami$^1$
}

\affiliation
{
$^1$Department of Physics, Kyoto University,
Kyoto 606-8502, Japan\\
$^2$Department of Applied Physics, Osaka University, 
Suita, Osaka 565-0871, Japan
}

\date{\today}

\begin{abstract}
We study ultracold fermionic atoms trapped in an optical lattice 
with harmonic confinement by combining the real-space 
dynamical mean-field theory with a two-site impurity solver. 
By calculating the local particle density and the pair potential
in the systems with different clusters,
we discuss the stability of a supersolid state, 
where an $s$-wave superfluid coexists with a density-wave state 
of checkerboard pattern.
It is clarified that a confining potential plays an essential role
in stabilizing the supersolid state. 
The phase diagrams are obtained for several effective particle densities.
\end{abstract}

\pacs{Valid PACS appear here}%

\maketitle

\section{Introduction}

Since the successful realization of Bose-Einstein condensation 
in a bosonic $\rm ^{87}Rb$ system~\cite{Rb}, 
ultracold atomic systems have 
attracted considerable interest.~\cite{Review,PethickSmith,Pitaevskii}
One of the most active topics in this field is an optical lattice system, 
\cite{BlochGreiner,Bloch,Jaksch,Morsch}
which is formed by loading the ultracold atoms 
in a periodic potential.
This provides a clean system with quantum parameters 
which can be tuned in a controlled fashion 
from weak to strong coupling limits.
In fact, remarkable phenomena have been observed 
such as the phase transition between a Mott insulator and a superfluid 
in bosonic systems~\cite{Greiner}. 
In addition, the superfluid state~\cite{Chin} and 
the Mott insulating state~\cite{Joerdens,Schneider} have been observed 
in the fermionic optical lattices,
which stimulates theoretical investigations on the quantum states 
in the optical lattice systems.
Among them, the possibilty of the supersolid state 
has been discussed as one of the interesting problems in optical lattice systems.
The existence of the supersolid state was experimentally 
suggested in a bosonic $^4$He system,~\cite{Kim}
and was theoretically discussed 
in the strongly correlated systems
such as bosonic systems~\cite{SSboson}
and Bose-Fermi mixtures~\cite{SSmix}. 
As for fermionic systems,
it is known that 
a density wave (DW) state and an $s$-wave superfluid (SSF) state are degenerate 
in the half-filled attractive Hubbard model on the bipartite lattice 
except for one dimension, \cite{Scalettar,Freericks}
which means that the supersolid state might be realizable in principle. 
However, the degenerate ground states are unstable against perturbations. 
In fact, the hole doping immediately drives the system to a genuine SSF state.
Therefore, it is difficult to realize the supersolid state in the homogeneous 
bulk system.
By contrast, in the optical lattice, an additional confining potential 
makes the situation different. 
\cite{Pour}
In our previous paper,\cite{Koga} 
we studied the attractive Hubbard model on square lattice 
with harmonic potential to clarify that the supersolid state is 
indeed realized at low temperatures.
However, we were not able to systematically deal with large clusters
to discuss how the supersolid state depends on 
the particle density, the system size, etc.
This might be important for experimental observations of the supersolid state
in the optical lattice.

In this paper, we address this problem
by combining the real-space dynamical mean-field theory (R-DMFT) 
with a two-site impurity solver.
We then discuss how stable 
the DW, SSF and supersolid states are
in the optical lattice system.
We also clarify the role of the confining potential 
in stabilizing the supersolid state.

The paper is organized as follows. 
In Sec. \ref{2}, we introduce the model Hamiltonian and explain 
the detail of R-DMFT and its impurity solver.
We demonstrate that the supersolid state is indeed realized 
in a fermionic optical lattice with attractive interactions in Sec. \ref{3}. 
In Sec. \ref{4}, we discuss the stability of the supersolid state 
in large clusters.
We also examine how the phase diagram depends on
the particle number.
A brief summary is given in Sec. \ref{5}.

\section{Model Hamiltonian and Method}\label{2}
Let us consider ultracold fermionic atoms in the optical lattice
with confinement, 
which may be described by the following attractive Hubbard model,
\cite{Scalettar,Garg,Capone,Micnas,Keller,Freericks,Toschi}
\begin{eqnarray}
H=-t\sum_{\langle ij \rangle \sigma} c_{i\sigma}^\dag c_{j\sigma}
-U\sum_i n_{i\uparrow}n_{i\downarrow} 
+ \sum_{i\sigma} v(r_i) n_{i\sigma},
\end{eqnarray}
where $c_{i\sigma} (c_{i\sigma}^\dag)$ annihilates (creates) 
a fermion at the $i$th site with spin $\sigma$ and 
$n_{i\sigma} = c_{i\sigma}^\dag c_{i\sigma}$.
$t(>0)$ is a nearest neighbor hopping, $U(>0)$ an attractive interaction, 
$v(r)\;[=V (r/a)^2]$ a harmonic potential and 
the term $\langle ij \rangle$ indicates that 
the sum is restricted to nearest neighbors.
$r_i$ is a distance measured from the center of the system and 
$a$ is lattice spacing.
Here, we define the characteristic length of the harmonic potential as
$d=(V/t)^{-1/2}a$, which satisfies the condition $v(d)=t$.

The ground-state properties of the Hubbard model 
on inhomogeneous lattices have theoretically been studied 
by various methods such as the Bogoljubov-de Gennes equations~\cite{FFLO}, 
the Gutzwiller approximation~\cite{Yamashita},
the slave-boson mean-field approach~\cite{Ruegg}, 
variational Monte Carlo simulations~\cite{Fujihara},
local density approximation.\cite{Dao}
Although magnetically ordered and superfluid states are 
described properly in these approaches, 
it may be difficult to describe the coexisting phase like a supersolid state 
in the inhomogeneous system.
The density matrix renormalization group 
method~\cite{Machida} and the quantum Monte Carlo method~\cite{QMC} 
are efficient for one-dimensional systems, but it may be difficult to 
apply them to higher dimensional systems with large clusters.
We here use R-DMFT~\cite{DMFT}, 
where local particle correlations are taken into account precisely. 
This treatment is formally exact for the homogeneous lattice model 
in infinite dimensions\cite{DMFT} and
the method has successfully been applied to some inhomogeneous 
correlated systems such as the surface~\cite{Potthoff} or 
the interface of the Mott insulators~\cite{Okamoto}, 
the repulsive fermionic atoms~\cite{Helmes,Snoek}. 
Furthermore, it has an advantage in treating 
the SSF state and the DW state on an equal footing in the strong coupling regime,
which allows us to discuss the supersolid state in the optical lattice. 

In R-DMFT, the lattice model is mapped to an effective impurity model, 
where local electron correlations are taken into account precisely. 
The lattice Green function is then obtained via self-consistent
conditions imposed on the impurity problem.
When one describes the superfluid state in the framework of R-DMFT,\cite{DMFT}
the lattice Green's function for the system size $L$ should be represented 
in the Nambu-Gor'kov formalism. 
It is explicitly given 
by the $(2L\times 2L)$ matrix,
\begin{widetext}
\begin{eqnarray}
\left[\hat{G}_{lat}^{-1}\left(i\omega_n \right)\right]_{ij}=
-t\delta_{\langle ij \rangle}\hat{\sigma}_z
+\delta_{ij}\left[ i\omega_n \hat{\sigma}_0 
+ \left\{\mu-v\left( r_i\right) \right\}\hat{\sigma}_z 
-\hat{\Sigma}_i\left(i\omega_n\right)\right],
\label{eq:lat}\end{eqnarray}
\end{widetext}
where $\hat{\bf \sigma}_\alpha \; (\alpha=x,y,z)$ is 
the $\alpha$th component of the $(2\times 2)$ Pauli matrix, 
$\hat{\sigma}_0$ the identity matrix, $\mu$ the chemical potential, 
$\omega_n = (2n+1)\pi T$ the Matsubara frequency, and $T$ the temperature. 
The site-diagonal self-energy at $i$th site is given 
by the following $(2\times 2)$ matrix,
\begin{equation}
\hat{\Sigma}_i\left(i\omega_n\right)=\left(
\begin{array}{cc}
\Sigma_i\left(i\omega_n\right) & S_i\left(i\omega_n\right)\\
S_i\left(i\omega_n\right) & -\Sigma_i^*\left(i\omega_n\right)
\end{array}
\right),
\end{equation}
where $\Sigma_i(i\omega_n) \; [S_i(i\omega_n)]$ 
is the normal (anomalous) part of 
the self-energy.
In R-DMFT, the self-energy at the $i$th site is obtained by solving 
the effective impurity model, 
which is explicitly given by the following Anderson Hamiltonian,
\cite{Garg,Toschi}
\begin{eqnarray}
H_{imp,i}&=&\sum_{k\sigma} E_{ik}
a_{ik\sigma}^\dag a_{ik\sigma}
+\sum_k
\left(D_{ik} a_{ik\uparrow} a_{ik\downarrow}+h.c.\right)
\nonumber\\
&+&\sum_{k\sigma}V_{ik}\left(c_{i\sigma}^\dag a_{ik\sigma}+
a_{ik\sigma}^\dag c_{i\sigma}\right)
\nonumber\\
&+&\epsilon_i\sum_{\sigma} c_{i\sigma}^\dag c_{i\sigma}- U c_{i\uparrow}^\dag c_{i\uparrow} c_{i\downarrow}^\dag c_{i\downarrow},\label{eq:imp}
\end{eqnarray}
where $a_{ik\sigma} (a_{ik\sigma}^\dag)$ annihilates (creates)
a fermion with spin $\sigma$ in the effective bath
and $\epsilon_i$ is the impurity level. 
We have here introduced the effective parameters in the impurity model
such as the spectrum of host particles $E_{ik}$, 
the pair potential $D_{ik}$ and
the hybridization $V_{ik}$. 
By solving the effective impurity model eq. (\ref{eq:imp}) for each site,
we obtain the site-diagonal self-energy and the local Green's function.
The R-DMFT self-consistent loop of calculations is iterated 
under the condition that the site-diagonal component of 
the lattice Green's function is equal to the local Green's function 
obtained from the effective impurity model as
$\left[\hat{G}_{lat}\left(i\omega_n\right)\right]_{ii}
=\hat{G}_{imp,i}\left(i\omega_n\right)$.

When R-DMFT is applied to our inhomogeneous system, 
it is necessary to solve the effective impurity models $L$ times by iteration.
Therefore, numerically powerful methods such as
quantum Monte Carlo simulations, the exact diagonalization method, 
and the numerical renormalization group method 
may not be efficient since they require long time to 
perform R-DMFT calculations. 
In this paper, we use a two-site approximation~\cite{2site,Higashiyama}, 
where the effective bath is replaced by only one site.
In spite of this simplicity, it has an advantage in taking into account 
both low- and high-energy properties reasonably well 
within restricted numerical resources~\cite{Higashiyama,Okamoto}.

In the two-site approximation, 
a non-interacting Green's function for the impurity model
at the $i$th site is simplified as,
\begin{eqnarray}
\left[\hat{\cal G}^0_{imp,i}\left(i\omega_n\right)\right]^{-1}&=&
i\omega_n \hat{\sigma}_0-\epsilon_i\hat{\sigma}_z\nonumber\\
&-&V_i \hat{\sigma}_z \frac{1}
{i\omega_n \hat{\sigma}_0-E_i \hat{\sigma}_z-D_i \hat{\sigma}_x} 
V_i\hat{\sigma}_z,
\end{eqnarray}
where the index $k$ was omitted.
The effective parameters $\{ E_i, D_i, V_i, \epsilon_i \}$ 
should be determined self-consistently so that
the obtained results properly reproduce the original lattice problem.
Here, we use the following equations,
\begin{eqnarray}
\epsilon_i&=&-\left.{\rm Re}
\left[\hat{\cal G}^{0}_{imp,i}\left(i\omega_n\right)\right]^{-1}
_{11}\right|_{n\rightarrow \infty}\\
V_i&=&\sqrt{\left(a-1\right)\left(\pi^2 T^2+E_i^2+D_i^2\right)}\\
D_i&=&\frac{b}{1-a},
\end{eqnarray} 
where $a={\rm Im} [\hat{\cal G}_i^{0}(i\omega_0)]^{-1}_{11}/\pi T$ and 
$b={\rm Re}[\hat{\cal G}_i^{0}(i\omega_0)]^{-1}_{12}$.
Furthermore, the number of particles is fixed in the non-interacting Green's
function, as
\begin{eqnarray}
n_0^{(i)}&=&2T \sum_{n=0}{\rm Re}\left[\hat{\cal G}_{imp,i}^{0}(i\omega_n)\right]_{11}
+\frac{1}{2}.
\end{eqnarray}
We can determine the effective parameters
$\{ E_i, D_i, V_i, \epsilon_i \}$ in terms of these equations.

Here, the effective particle density is defined as $\tilde{\rho}=N/\pi d^2$,
where $N$ is the total number of particles. 
This density relates the systems with different sites, number of particles, 
and curvatures of the confining potentials 
in the same way as the particle density does for periodic systems 
with different sites.
We set $t$ as a unit of energy and calculate the density profile 
$\langle n_{i\sigma} \rangle = 2T\sum_{n=0}{\rm Re}[
 G_{i\sigma}\left(i\omega_n\right)]+\frac{1}{2}$ and the distribution of
the pair potential
$\Delta_i = 2T\sum_{n=0}{\rm Re} [F_i\left(i\omega_n\right)]$, 
where $G_{i\sigma}(i\omega_n) [F_i(i\omega_n)]$ 
is the normal (anomalous) Green's function for the $i$th site.  
Note that $\Delta_i$ represents the order parameter for the SSF state
for the $i$th site.

In the following, we consider the attractive Hubbard model on square lattice 
with harmonic confinement 
as a simple model for the supersolid. 
In this case, it is known that the symmetry of the square lattice 
is not broken in the DW, SSF, and supersolid states~\cite{Machida,FFLO,Koga}. 
Therefore, the point group $C_{4v}$ is useful to 
deal with the system on the inhomogeneous lattice.
For example, when the system with 5513 sites $(r<42.0)$ is treated, 
one can deal with only 725 inequivalent sites.
This allows us to discuss the low temperature properties 
in larger clusters,
in comparison with those with $(d/a=6.5, N\sim 300)$ 
treated in our previous paper.

\begin{figure}[tb]
\vskip -2mm
\begin{center}
\includegraphics[width=7cm]{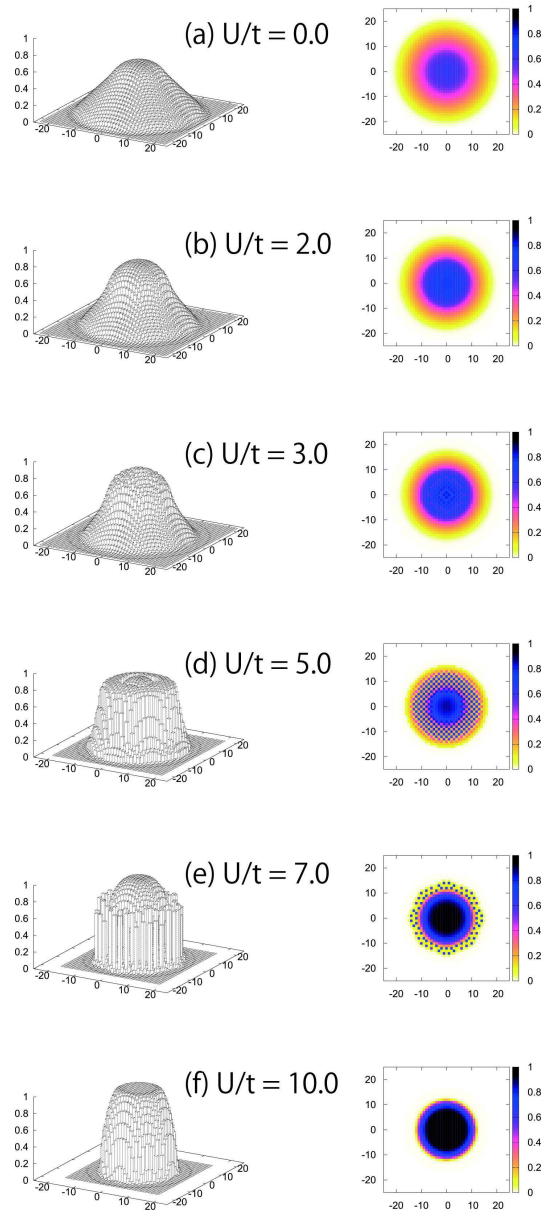}
\end{center}
\vskip -6mm
\caption{
(Color online) The density profile $\langle n_{i\sigma}\rangle$ in the optical lattice system 
with $d/a=10$ at $T/t=0.05$ when $U/t = 0.0, 2.0, 3.0, 5.0, 7.0$ and $10.0$
(from the top to the bottom).
}
\label{fig:fig1}
\end{figure}
\begin{figure}[tb]
\vskip -2mm
\begin{center}
\includegraphics[width=7cm]{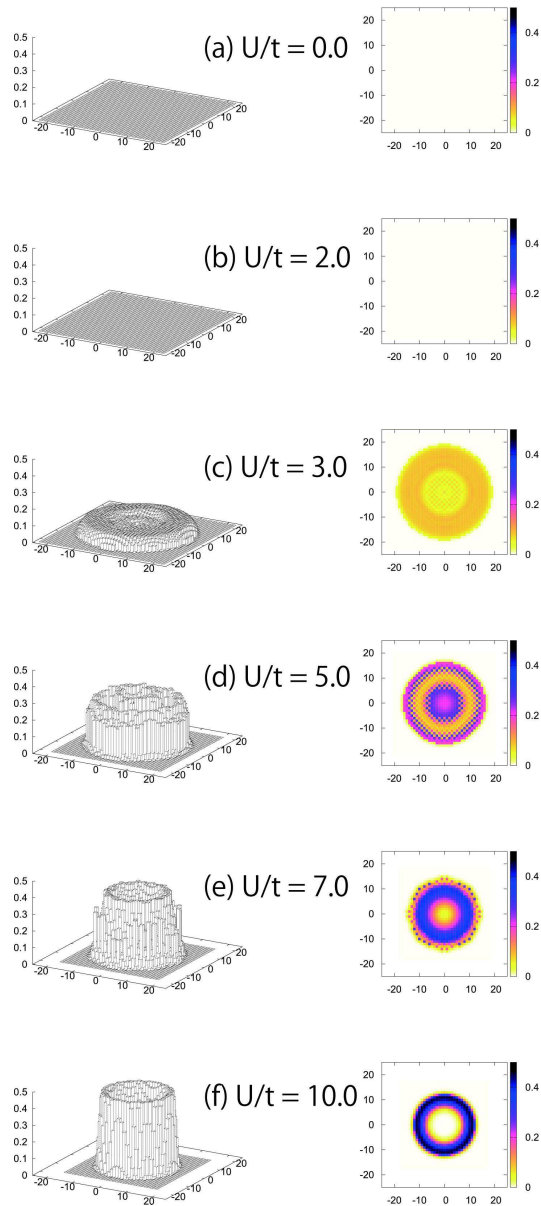}
\end{center}
\vskip -6mm
\caption{
(Color online) The pair potential $\Delta_i$ in the optical lattice system 
with $d/a=10$ at $T/t=0.05$ when $U/t = 0.0, 2.0, 3.0, 5.0, 7.0$ and $10.0$
(from the top to the bottom).
}
\label{fig:fig2}
\end{figure}

\section{Low temperature properties}\label{3}
By means of R-DMFT with the two-site impurity solver, 
we obtain the results for the system with $d/a=10$ and 
$N\sim 720 (\tilde{\rho}\sim 2.3)$.
Figures \ref{fig:fig1} and \ref{fig:fig2} show 
the profiles of the local density and the pair potential at $T/t=0.05$.
In the non-interacting case $(U/t=0)$, 
fermionic atoms are smoothly distributed up to $r/a\sim 21$,
as shown in Fig. \ref{fig:fig1} (a).
Increasing the attractive interaction $U$, 
fermions tend to gather around the bottom of the harmonic potential, 
as seen in Fig. \ref{fig:fig1} (b). 
In these cases, the pair potential is not yet developed, as shown 
in Figs. \ref{fig:fig2} (a) and (b), and thereby
the normal metallic state with short-range pair correlations 
emerges in the region ($U/t \lesssim 2$).  
Further increase in the interaction $U$ leads to different behavior, where 
the pair potential $\Delta_i$ is induced
in the region with $\langle n_{i\sigma} \rangle \neq 0$.
Thus, the SSF state is induced by the attractive interaction, 
which is consistent with the results obtained from 
the Bogoljubov-de Gennes equation~\cite{FFLO}. 
In the case with $U/t=3$, 
another remarkable feature is found around the center of 
the harmonic potential $(r/a<7)$, where a checkerboard structure appears 
in the density profile $\langle n_{i\sigma} \rangle$,
as shown in Fig. \ref{fig:fig1} (c).
This implies that the DW state is realized in the region.
On the other hand, the pair potential $\Delta_i$ is not suppressed completely 
even in the DW region, as shown in Fig. \ref{fig:fig2} (c).
This suggests that 
the DW state coexists with the SSF state, 
{\it i.e.}  {\it a supersolid state} appears in our optical lattice system. 
The profile characteristic of the supersolid state is clearly seen 
in the case of $U/t=5$.
Figures \ref{fig:fig1}(d) and \ref{fig:fig2}(d) show that
the DW state of checkerboard structure coexists with the SSF state 
in the doughnut-like region $(5<r/a<15)$.
By contrast, the genuine SSF state appears inside and outside of
the region $(r/a<5, 15<r/a<17)$.
Further increase in the interaction excludes the DW state out of the center
since fermionic atoms are concentrated around the bottom of the potential 
for large $U$.  
In the region, two particles with opposite spins are strongly coupled 
by the attractive interaction to form a hard-core boson, 
giving rise to the band insulator with $\langle n_{i\sigma}\rangle\sim 1$,
instead of the SSF state. 
Therefore, the SSF state survives only in the narrow circular region 
surrounded among
the empty and fully occupied states.
We see such behavior more clearly in Figs. \ref{fig:fig1} (f) and 
\ref{fig:fig2} (f). 
Note that in the strong coupling limit $U/t\rightarrow \infty$,
all particles are condensed in the region 
$r<r_{c}=\sqrt{\tilde{\rho}/2}\;d \sim 1.07d = 10.7a$.

In this section, we have studied the attractive Hubbard model 
with the harmonic potential to clarify that the supersolid state
is realized in a certain parameter region.
However, it is not clear how the supersolid state depends on the system size
and the number of particles.
To make this point clear,
we deal with large clusters to clarify
that the supersolid state is indeed realized in the following.

\section{Stability of the supersolid state}\label{4}

\begin{figure}[htb]
\begin{center}
\includegraphics[width=7cm]{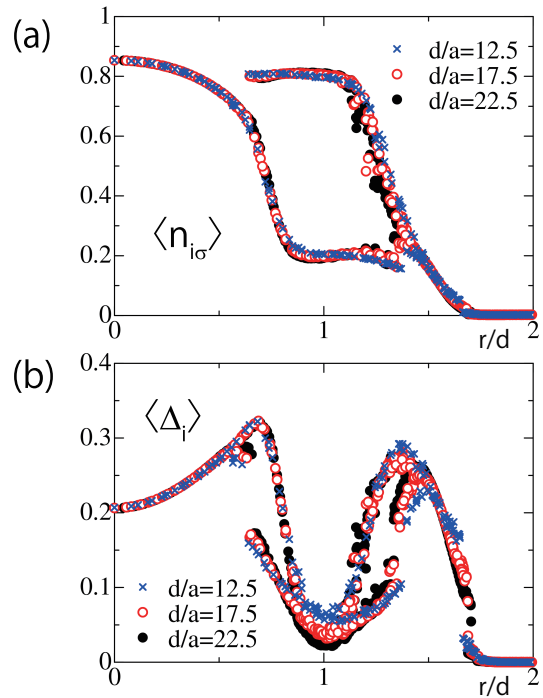}
\end{center}
\caption{
Profiles of particle density $\langle n_{i\sigma} \rangle$ 
and pair potential $\Delta_i$
as a function of $r/d$ with fixed $d=12.5, 17.5$ and $22.5$,
when $U/t=5$.
}
\label{fig:thermodynamic}
\end{figure}

In this section, we discuss the stability of the supersolid state 
in fermionic optical lattice systems,
which may be important for experimental observations.
First,
we clarify how low-temperature properties depend on the system size,
by performing R-DMFT for several clusters with different $d$.
We here fix $U/t=5$ and $\mu/t\sim -1.58$ to obtain the profiles of 
the local particle density and 
the pair potential, which are shown in Fig. \ref{fig:thermodynamic}.
We note that the distance $r$ is normalized by $d$ in the figure.
It is found that $\langle n_{i\sigma}\rangle$ and $\Delta_i$ 
describe smooth curves 
for $r/d \lesssim 0.6$ and $1.4 \lesssim r/d \lesssim 1.7$, 
where the genuine SSF state is realized. On the other hand,
for $0.6 \lesssim r/d \lesssim 1.4$, two distinct magnitudes appear 
in $\langle n_{i\sigma}\rangle$, 
reflecting the fact that the DW state with two sublattices is realized.
Since the pair potential is also finite in the region, 
the supersolid state is realized.
In this case, we deal with finite systems, and thereby
all data are discrete in $r$. 
Nevertheless, it is found that the obtained results are well scaled by $d$
although some fluctuations appear due to finite-size effects
in the small $d$ case.
The effective particle density $\tilde{\rho}$ 
is almost constant in the above cases. 
Therefore, we conclude that 
when $\tilde{\rho} \sim 2.3$,
the supersolid state discussed here is stable 
in the limit with $N, d\rightarrow \infty$.
This result does not imply that the supersolid state
is realized in the homogeneous system with arbitrary fillings.
In fact, the supersolid state might be realizable 
only at half filling~\cite{Scalettar,Freericks,Capone,Garg,Micnas}.
Therefore, we can say that a confining potential is essential to stabilize 
the supersolid state in the optical lattice system.

Next, we focus on the system with $U/t=5$ and $d/a=10$ to discuss in detail
how the supersolid state depends on
the effective particle density $\tilde{\rho}$.
The DW state is characterized by the checkerboard structure 
in the density profile $\langle n_{i\sigma} \rangle$, so that
the Fourier transform $n_q$ at $q=(\pi, \pi)$ is appropriate 
to characterize the existence of the DW state.
On the other hand, the Fourier transform $\Delta_q$ at $q=(0,0)$ 
may represent the rigidity of the SSF state in the system.
\begin{figure}[htb]
\begin{center}
\includegraphics[width=7cm]{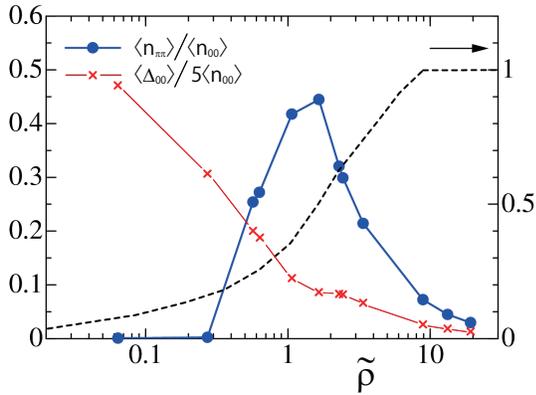}
\end{center}
\caption{
(Color online) $\langle n_{\pi\pi} \rangle/\langle n_{00}\rangle$
and $\Delta_{00}/5\langle n_{00} \rangle$
as a function of the effective particle density $\tilde{\rho}(=N/\pi d^2)$
when $U=5t$ and $T=0.05t$. A broken line represents
the local particle density at the center of the lattice 
in the noninteracting case.
}
\label{fig:rho}
\end{figure}
In Fig. \ref{fig:rho}, we show the semilog plots of 
the parameters normalized by $\langle n_{00} \rangle$.
It is found that the normalized parameter $\Delta_{00}$ is always finite
although the increase in the attractive interaction monotonically decreases it. 
This implies that the SSF state appears in the system 
with the arbitrary particle number.
In contrast to this SSF state, 
the DW state is sensitive to the effective particle density
as shown in Fig. \ref{fig:rho}.
These may be explained by the fact that 
in the system without a harmonic confinement $(V_0=0)$,
the DW state is realized only at half filling $(n=0.5)$,
while the SSF state is always realized.
To clarify this, we also show the local particle density 
in the noninteracting case 
at the center of the system as the broken line in Fig. \ref{fig:rho}.
It is found that when the quantity approaches 
half-filling $(\sim 0.5)$, $\langle n_{\pi\pi}\rangle/\langle n_{00}\rangle$ 
takes its maximum value, where the SSF state coexists with the DW state. 
Therefore, we can say that the supersolid state is stable around this condition.
Increasing the effective particle density,
the band insulating states become spread around the center, while
the DW and SSF states should be realized 
in a certain circular region surrounded among the empty and 
fully-occupied regions.
Therefore, the normalized parameters 
$\langle n_{\pi\pi}\rangle$ and $\Delta_{00}$ 
are decreased with increase in $\tilde{\rho}$.
On the other hand, in the case with low density $\tilde{\rho}\lesssim 0.3$, 
the local particle density $n_i$ at each site is far from half filling 
even when $U/t=5$.
Therefore, the DW state does not appear in the system, but 
the genuine SSF state is realized.
These facts imply that the condition $n\sim0.5$ is still important 
to stabilize the supersolid state even in fermionic systems
confined by a harmonic potential.\cite{Pour}

\begin{figure}[htb]
\begin{center}
\includegraphics[width=7cm]{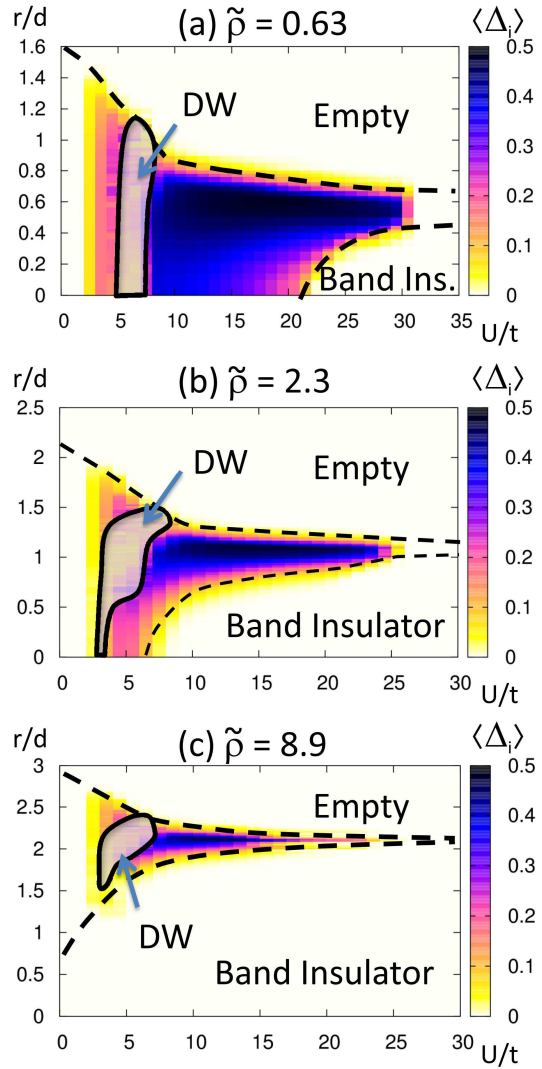}
\end{center}
\caption{
(Color online) The phase diagram of the attractive Hubbard model 
on the optical lattice 
with $\tilde{\rho}\sim 0.63, 2.3$ and $8.9$.
The density plot represents the profiles of the $s$-wave pair potential
as a function of the attractive interaction $U/t$. 
The DW state is realized in the shaded area. 
The broken lines give a guide to eyes which distinguishes the region 
with a fractional particle density from the empty 
and fully-occupied regions.
}
\label{fig:phases}
\end{figure}

By performing similar calculations for the systems with 
low, intermediate, and high particle densities 
($\tilde{\rho}\sim 0.63, 2.3$ and $8.9$), 
we end up with the phase diagrams, as shown in Fig. \ref{fig:phases}.
We find that increasing the attractive interaction,
fermionic particles gradually gather around the center of the system,
where the empty state is stabilized away from the center and 
the band insulating state with fully occupied sites 
is stabilized.
It is found that the region surrounded among these states strongly depend 
on the effective particle density $\tilde{\rho}$. 
The increase in the effective particle density shrinks
the region, 
which affects the stability of the SSF, DW and their coexisting states. 
In particular, the DW region, 
which is shown as the shaded area in Fig. \ref{fig:phases},
is sensitive to the effective particle density, as discussed above.
Namely, the local pair potential $\Delta_i$ takes its maximum value around 
$U/t \sim 15$, 
which may give a rough guide for the crossover region between
the BCS-type and the BEC-type states.
We note that the DW state appears only in the BCS region
$(U/t\sim 5)$.
This implies that the condition $n\sim 0.5$ is not sufficient, 
but necessary to stabilize the supersolid state
in the attractive Hubbard model with an inhomogeneous potential. 

We wish to comment on the conditions to observe the supersolid state 
in the fermionic optical lattice system.
Needless to say, one of the most important conditions 
is the low temperature.\cite{Koga}
Second is the tuning of
the effective particle density $\tilde{\rho} (\sim 1)$, 
which depends on the curvature of 
the harmonic potential as well as the total number of particles.
This implies that 
a confined potential play a crucial role in stabilizing 
the supersolid state in fermionic optical lattice systems.
In addition to this, an appropriate attractive interaction is necessary 
to stabilize the DW state in the BCS-type SSF state.
When these conditions are satisfied, the supersolid state is expected 
to be realized at low temperatures.

\section{Summary}\label{5}
We have investigated the fermionic attractive Hubbard model 
in the optical lattice with harmonic confinement. 
By combining R-DMFT with a two-site impurity solver, 
we have obtained the rich phase diagram on the square lattice, 
which has a remarkable domain structure 
including the SSF state in the wide parameter region. 
By performing systematic calculations,
we have then confirmed that the supersolid state, 
where the SSF state coexists with the DW state, 
is stabilized even in the limit with 
$N\rightarrow \infty, V_0 \rightarrow 0$ and $\tilde{\rho}\sim$ const. 
We have also elucidated that a confining potential plays a key role 
in stabilizing the supersolid state. 

\acknowledgments
This work was supported by the Grant-in-Aid for the Global COE Program 
"The Next Generation of Physics, Spun from Universality and Emergence" 
and Scientific Research 
[20740194 (A.K.), 20540390 (S.S.), 19014013 and 20029013 (N.K.)]
from the Ministry of Education, Culture, Sports, Science and Technology (MEXT) 
of Japan. 
Some of the computations were performed at the Supercomputer Center at the 
Institute for Solid State Physics, University of Tokyo.

%

\end{document}